\documentclass{article}
\usepackage[utf8]{inputenc}
\usepackage{multicol,caption}
\usepackage{graphicx}
\usepackage[margin=2cm]{geometry}
\usepackage{titlesec}
\usepackage{setspace}
\usepackage{hyperref}
\usepackage{amsmath}
\usepackage{caption}
\usepackage{subcaption}
\usepackage{authblk}
\usepackage{array}
\usepackage{float}
\usepackage{appendix}
\usepackage{xcolor}
\usepackage{url}
\usepackage{multirow}
\definecolor{cyan(process)}{rgb}{0.0, 0.72, 0.92}
\definecolor{very_light_gray}{gray}{0.85}
\definecolor{mygray}{gray}{0.85}

\titleformat{\section}[block]{\normalsize\bfseries\filcenter}{}{1em}{}
\titlespacing*{\section}{0pt}{11pt plus 2pt minus 2pt}{11pt plus 2pt minus 2pt}

\titleformat{\subsection}{\normalsize\itshape\filcenter}{}{1em}{}
\titlespacing*{\subsection}{0pt}{11pt plus 2pt minus 2pt}{11pt plus 2pt minus 2pt}
\newcommand{\blockcomment}[1]{}

\newenvironment{Figure}
  {\par\medskip\noindent\minipage{\linewidth}}
  {\endminipage\par\medskip}

  \usepackage{tikz}
  \usetikzlibrary{decorations.pathreplacing, calc}
  \usetikzlibrary{shapes.geometric, arrows}
  \tikzstyle{startstop} = [
                  rectangle, 
                  rounded corners, 
                  minimum width=3cm, 
                  minimum height=1cm,
                  text centered, 
                  draw=black, 
                  fill=red!30
                  ]
  \tikzstyle{io} = [
                  trapezium, 
                  trapezium stretches=true,
                  trapezium left angle=70, 
                  trapezium right angle=110, 
                  minimum width=1cm, 
                  minimum height=0.7cm, 
                  text centered, 
                  text width=2.5cm, 
                  draw=black, 
                  fill=blue!30
                  ]
  \tikzstyle{process} = [
                  rectangle, 
                  minimum width=2.5cm, 
                  minimum height=0.7cm, 
                  text centered, 
                  text width=3.0cm, 
                  draw=black, 
                  fill=orange!30
                  ]
    \tikzstyle{process2} = [
                  rectangle, 
                  minimum width=3cm, 
                  minimum height=0.7cm, 
                  text centered, 
                  text width=2.2cm, 
                  draw=black, 
                  fill=orange!30
                  ]
  \tikzstyle{decision} = [
                  diamond, 
                  minimum width=1.3cm, 
                  minimum height=1cm, 
                  text centered, 
                  text width=1.3cm, 
                  draw=black, 
                  fill=mygray
                  ]
  \tikzstyle{arrow} = [
                  thick,->,>=stealth
                  ]

\title{\textbf{Optimisation of Pulse Waveforms for Qubit Gates using Deep Learning}}
\author[1]{Zachary Fillingham}
\author[1]{Hossein Nevisi}
\author[1]{Shirin Dora \thanks{Corresponding author: S.Dora@lboro.ac.uk}}
\affil[1]{Department of Computer Science, Loughborough University}
\date{}

\begin{document}
\maketitle

\vspace{-0.5cm}
\begin{abstract}
    In this paper, we propose a novel method using Deep Neural Networks (DNNs) to optimise the parameters of pulse waveforms used for manipulating qubit states, resulting in high fidelity implementation of qubit gates.\blockcomment{we will be delving into the very frontier of the interdisciplinary field of quantum computing and applications of deep learning. This is an extremely new area of research, but incredibly important since it could allow for} High fidelity quantum simulations are crucial for scaling up current quantum computers. The proposed approach uses DNNs to model the functional relationship between amplitudes of pulse waveforms used in scheduling and the corresponding fidelities. The DNNs are trained using a dataset of amplitude and corresponding fidelities obtained through quantum simulations in Qiskit. A two-stage approach is used with the trained DNNs to obtain amplitudes that yield the highest fidelity. The proposed method is evaluated by estimating the amplitude for pulse scheduling of single (Hadamard and Pauli-X) and two qubit gates (CNOT). The results clearly indicate that the method can achieve high fidelity implementations of single-qubit gates with fidelities of 0.999976 and 0.999923 for Hadamard and Pauli-X gates, respectively. For the CNOT gate, the best fidelity obtained is 0.695313. This can be attributed to the effects of entanglement and the need for the phase parameter to be accounted for within the predictive model. 
  \end{abstract}

\begin{multicols}{2}
\section{I. INTRODUCTION}
\label{secIIB}
Quantum computing has emerged as a promising technology in the realm of computational science, with the potential processing power well beyond the boundaries of classical computing. In an era where complex problems continue to challenge the limits of classical computers, quantum computing offers a transformative approach to computation. 

Quantum computing harnesses the fundamental principles of quantum mechanics. At its core are quantum bits, or qubits, the quantum analog of classical bits. Qubits can exist in a superposition of two base states, representing both 0 and 1 simultaneously. 
This inherent property enables quantum systems to process information in parallel, leading to exponentially more efficient computations than classical systems for certain problems \cite{divincenzo1995quantum}.

The superior processing power of quantum computers gives them the potential to solve problems that have long confounded classical machines, like cryptography \cite{aumasson2017impact}, optimisation of material science \cite{bauer2020quantum}, drug discovery \cite{cao2018potential}. In addition to superposition, qubits exhibit entanglement, which links two states even when they are far apart \cite{monroe2016quantum}, enhancing the correlation between qubits. This enables the changing of the state of an entangled qubit instantaneously changing the state of the paired qubit in quantum computers, leading to accelerated processing speeds.

Within quantum computing there are three main types of qubits \cite{sanders2017build}: trapped ions with long coherence times but scalability challenges \cite{gerritsma2011quantum}, photonic qubits suitable for communication \cite{sanders2017build}, and superconducting qubits which are the focus of this paper. Superconducting qubits have enabled development of complex circuits with large number of qubits \cite{chen2018metrology}. Similar to classical computers, quantum circuits rely on fundamental qubit gates, such as the Hadamard and Pauli-X gates, to perform computations by changing the states of qubits. Qubit gates manipulate the states of qubits using pulse waveforms, which are precisely tailored electromagnetic signals. In essence, the electromagnetic pulses act as control knobs that adjust the quantum states of the qubits, analogous to how electronic circuits use voltage signals to control electronic components.

Quantum computing's transformative potential is severely hindered by the susceptibility of qubits to noise, which induce errors in quantum circuits. Fidelity is a measure of the accuracy of a qubit gate's operation. It is a crucial metric that determines the reliability of quantum circuits. To achieve high fidelity quantum operations, pulse shaping techniques are used to find pulse waveforms that manipulate the qubits more precisely. Pulse shaping involves altering aspects of the electromagnetic pulses, such as their amplitude, duration, and phase, thereby shaping the pulse waveform to improve control over qubit states.

\begin{figure*}[t] 
  \centering
  \begin{tikzpicture}
  \draw[gray!30,fill=gray!10] (-8,-0.8) rectangle (-3.1,2.8);
  \draw[gray!30,fill=gray!10] (-2.6,-0.8) rectangle (2.3,2.8);
  \draw[gray!30,fill=gray!10] (2.6,-0.8) rectangle (7.7,2.8);
    \node [draw=gray!100, inner sep=8pt, line width=0.5pt, rounded corners=5pt] {
  \begin{subfigure}[b]{0.3\textwidth}
    \includegraphics[width=\linewidth]{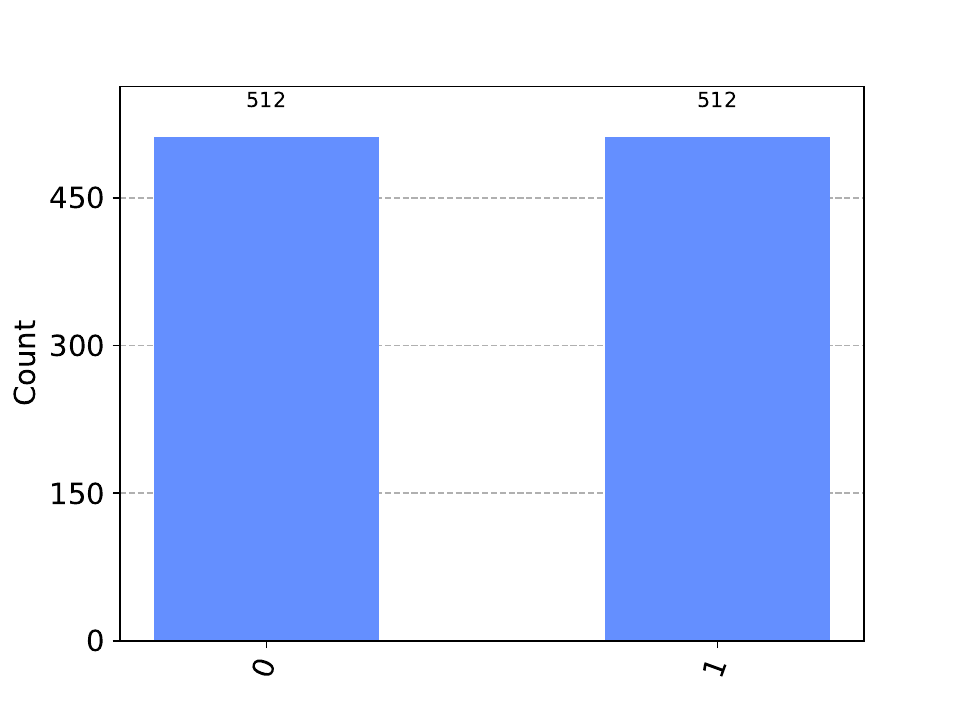}
    \caption{Histogram of the ideal equal distribution between the two states $|0\rangle$ and $|1\rangle$ after the Hadamard gate has been applied to a qubit initialised as $|0\rangle$: 50\% split between states.} 
    \label{hist_ideal_h}
  \end{subfigure}
  \hfill
  \begin{subfigure}[b]{0.3\textwidth}
    \includegraphics[width=\linewidth]{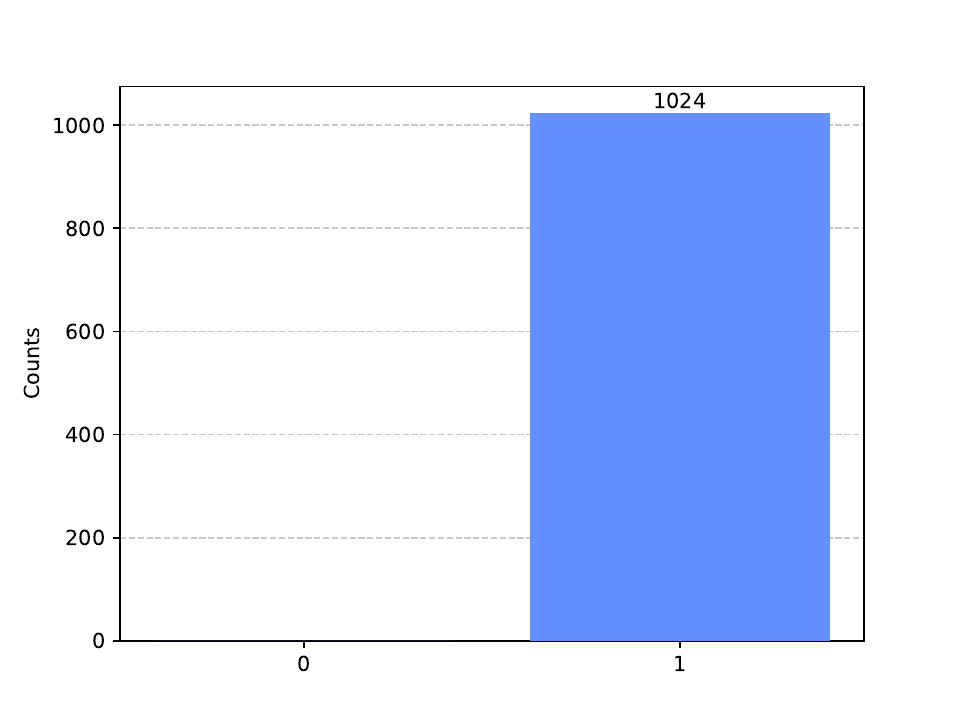}
    \caption{Histogram of the ideal distribution between the two states $|0\rangle$ and $|1\rangle$ after the Pauli-X gate has been applied to a qubit initialised as $|0\rangle$: 100\% probability of getting state $|1\rangle$.}
    \label{hist_ideal_x}
  \end{subfigure}
  \hfill
  \begin{subfigure}[b]{0.3\textwidth}
    \includegraphics[width=\linewidth]{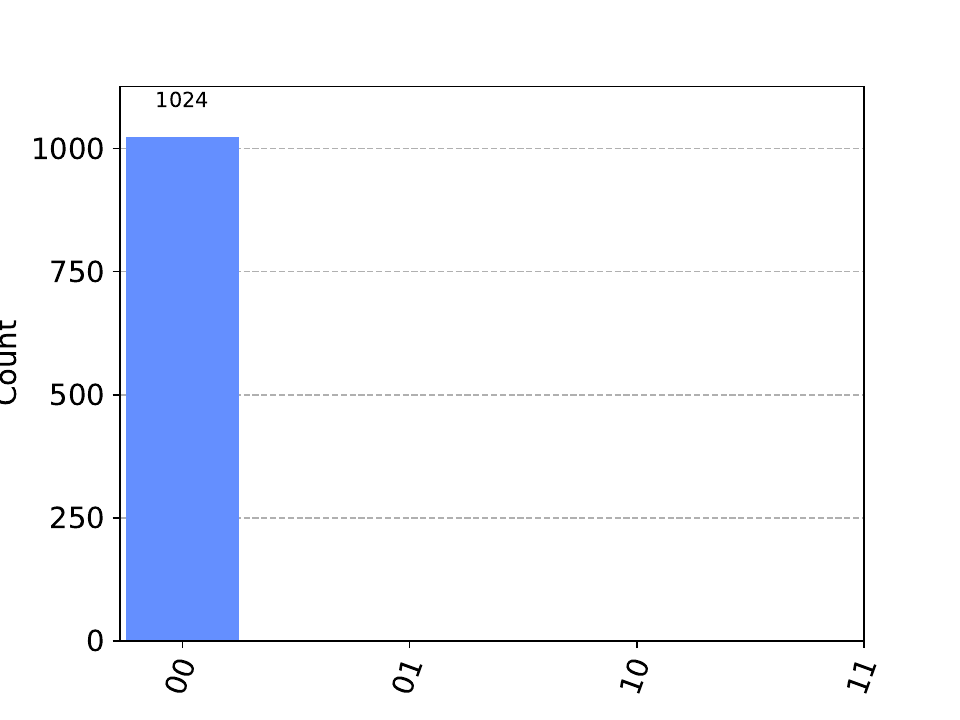}
    \caption{The ideal distribution between the states $|00\rangle$, $|01\rangle$, $|10\rangle$, and $|11\rangle$ after the CNOT gate has been applied to two qubits, both initialised as $|0\rangle$: state $|00\rangle$ has 100\% probability.}
    \label{hist_ideal_cnot}
  \end{subfigure}
  };
  \end{tikzpicture}
  \caption{\blockcomment{Theoretically}Ideal results for Hadamard and Pauli-X gates on a qubit initialized at $|0\rangle$, and for the CNOT gate on both qubits initialized at $|0\rangle$. The count represents the number of times each state was measured during the simulation.}
  \label{Ideal_distributions}
\end{figure*}

The L-BFGS-B optimisation algorithm \cite{zhu1997algorithm} has previously been employed for pulse shaping by iteratively refining the control pulses \cite{matekole2022methods} and enhance qubit gate fidelity. L-BFGS-B algorithm employs a physics-guided quantum optimal control (QOC) strategy. QOC strategies involve adjusting the shape and timing of the external electromagnetic fields to increase the efficiency of a quantum dynamical processes. L-BFGS-B minimizes a cost function representing gate infidelity in the search space of control pulses with different amplitudes. This method has improved qubit gate fidelity; however, this relies on generating pulses that are tested and refined iteratively through qubit gate simulations in Qiskit. L-BFGS-B is a computationally expensive approach due to iterative steps that involve quantum simulation of pulses with different amplitudes.

In this paper, we utilize deep neural networks (DNNs) do develop a computationally efficient solution for pulse shaping. DNNs can capture complex relationships within data, making them well-suited for this task. The intersection of deep learning and quantum computing is a relatively new field. Existing papers have mostly focussed on addressing circuit design and quantum control theory for qubit gate optimisation \cite{fosel2021quantum, niu2019universal, bukov2018reinforcement}. Here, DNNs are used for optimising the amplitudes of the pulse waveforms that control the qubit gates, by training them to predict the fidelities given a pulse amplitude.

The key idea in this paper is to train DNNs to learn the functional relationship between the amplitudes of the pulse waveform and the fidelity for each qubit gate. The training data containing amplitudes of pulse waveforms and fidelity is obtained through quantum simulations. 

A two-stage approach is used to obtain pulse waveform amplitudes with high fidelities. In the first stage, the trained neural network is used to obtain predictions of fidelity given random and coarsely selected values for pulse waveform amplitudes. Based on the predictions, the amplitudes corresponding to high fidelities are identified. In the second stage, a new set of amplitude values are sampled in the neighborhood of amplitudes with high fidelities identified in the first stage. The trained neural network is used to obtain predictions for the amplitudes sampled in the second stage. The amplitude with highest fidelity in this stage is used in quantum simulations for the specific qubit gate. This two-stage approach does not rely on quantum simulations to obtain high fidelity pulse waveforms.

The proposed method is evaluated for estimating amplitudes with high fidelity for Hadamard, Pauli-X and CNOT gates. The prediction results demonstrate that the proposed methods can achieve higher than 99\% fidelity for Hadamard and Pauli-X gates, and over 69\% for the CNOT gate.

Rest of the paper is organized as follows. Section ~\hyperref[background]{\textcolor{cyan(process)}{II}} presents the relevant theoretical concepts in this paper including qubit gates. Section~\hyperref[qiskit_pulse_sec]{\textcolor{cyan(process)}{III}} explains the pulse-level simulations used to obtain data for training the DNNs and the proposed method for identifying amplitudes resulting in high fidelities. Section~\hyperref[result_sec]{\textcolor{cyan(process)}{IV}} presents the results of using the proposed method.

\begin{figure*}[t] 
  \centering
  \begin{subfigure}[b]{\textwidth}
  \centering
    \begin{tikzpicture}
    
        \draw[gray!30,fill=gray!10] (-5.5,-6) rectangle (12,-0.5);

        \node at (-2.4,-1.2) {Pulse Amplitudes};
        
        \begin{scope}[shift={(-2.4,-3.5)}]
            \node at (0,0) {\includegraphics[width=5cm, height=3.5cm]{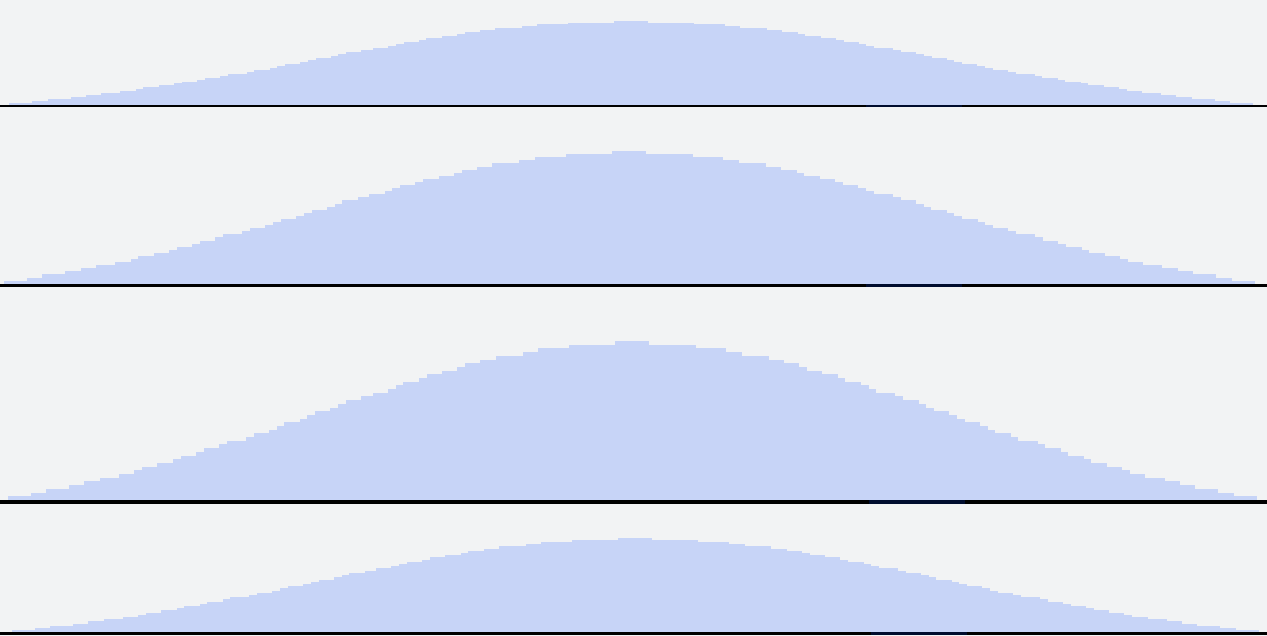}};
            \draw[black, ultra thin, |-|] (0, 1.17) -- (0, 1.64);
            \node at (0.25, 1.4) {\(\phi_1\)};
            
            \draw[black, ultra thin, |-|] (0, 0.19) -- (0, 0.91);
            \node at (0.25, 0.55) {\(\phi_2\)};
            
            \draw[black, ultra thin, |-|] (0, -0.14) -- (0, -1.0);
            \node at (0.25, -0.6) {\(\phi_3\)};
            
            \draw[black, ultra thin, |-|] (0, -1.72) -- (0, -1.22);
            \node at (0.25, -1.5) {\(\phi_4\)};
        \end{scope}

        \draw[line width=5pt, gray, -latex] (0.7,-4) -- (1.7,-4);

        \node at (3.5,-1.2) {Quantum Simulation};
        
        \begin{scope}[scale=1.5, shift={(0.9,-2.7)}]
            \draw (1.5,0) circle (1cm);
            \draw[->] (1.5,0) -- (2.8,0) node[right] {$y$};
            \draw[->] (1.5,0) -- (1.5,1.4) node[above] {$z$};
            \draw[->] (1.5,0) -- (0.95,-0.45) node[below] {$x$};
            \draw[->] (1.5,0) -- (2.5,1) node[above] {$|\psi\rangle $};
            \draw[dashed] (1.5,0) ellipse (1cm and 0.3cm);
            \filldraw (1.5,1) circle (2pt) node[right,xshift=0mm,yshift=2mm] {$|0\rangle$};
            \filldraw (1.5,-1) circle (2pt) node[right,xshift=0mm,yshift=-2mm] {$|1\rangle$};
        \end{scope}
    
        \draw[line width=5pt, gray, -latex] (6.3,-4) -- (7.3,-4);

        \node[align=center] at (9.1,-1.2) {Amplitude-Fidelity \\ Dataset};
        
        \begin{scope}[shift={(9.1,-3)}]
            
            \filldraw[fill=gray!30] (0,0) ellipse (1.2 and 0.5);
            \draw (-1.2,0) -- (-1.2,-2);
            \draw (1.2,0) -- (1.2,-2);
            \draw (-1.2,-2) arc (180:360:1.2 and 0.5);
            
            \node at (0,-1.0) {Amplitude: $\phi$};
            \node at (0,-1.7) {Fidelity: $\chi$};
        \end{scope}
        
    \end{tikzpicture}
    \caption{Schematic of data generation: Pulses of different amplitudes, $\phi$, are used within a quantum simulation. The quantum simulation can be any type of gate operation and is represented here by a Bloch sphere. This simulation yields the fidelity, $\chi$, by comparing the result to the theoretical ideal. This method is used to collect a dataset of amplitude-fidelity combinations \( (\phi_i, \chi_i) \) for each quantum gate.}
    \label{data_gen_schematic}
  \end{subfigure}
  \hfill
  \begin{subfigure}[b]{\textwidth}
  \centering
    \begin{tikzpicture}
        \draw[gray!30,fill=gray!10] (-5.5,-5.5) rectangle (12,0);

        \node at (-1.0,-0.4) {Deep Neural Network};
        
        \begin{scope}[x=0.6cm, y=0.8cm,shift={(-8.4,-4.3)}]
            \node[circle,draw,fill=red!20] (input) at (0,0) {};
            
            \node[circle,draw,fill=blue!20] (h11) at (2,0.75) {};
            \node[circle,draw,fill=blue!20] (h13) at (2,-0.75) {};
            \path (h11) -- (h13) node [black, font=\LARGE, midway, sloped, xshift=0.8mm] {$\dots$};
            
            \node[circle,draw,fill=blue!20] (h21) at (4,1.5) {};
            \node[circle,draw,fill=blue!20] (h23) at (4,0) {};
            \path (h21) -- (h23) node [black, font=\LARGE, midway, sloped, xshift=0.8mm] {$\dots$};
            \node[circle,draw,fill=blue!20] (h25) at (4,-1.5) {};
            \path (h23) -- (h25) node [black, font=\LARGE, midway, sloped, xshift=0.8mm] {$\dots$};
            
            \node[circle,draw,fill=blue!20] (h31) at (6,2.25) {};
            \node[circle,draw,fill=blue!20] (h33) at (6,0.75) {};
            \path (h31) -- (h33) node [black, font=\LARGE, midway, sloped, xshift=0.8mm] {$\dots$};
            \node[circle,draw,fill=blue!20] (h34) at (6,0) {};
            \node[circle,draw,fill=blue!20] (h35) at (6,-0.75) {};
            \node[circle,draw,fill=blue!20] (h37) at (6,-2.25) {};
            \path (h35) -- (h37) node [black, font=\LARGE, midway, sloped, xshift=0.8mm] {$\dots$};
            
            \node[circle,draw,fill=blue!20] (h41) at (8,1.5) {};
            \node[circle,draw,fill=blue!20] (h43) at (8,0) {};
            \path (h41) -- (h43) node [black, font=\LARGE, midway, sloped, xshift=0.8mm] {$\dots$};
            \node[circle,draw,fill=blue!20] (h45) at (8,-1.5) {};
            \path (h43) -- (h45) node [black, font=\LARGE, midway, sloped, xshift=0.8mm] {$\dots$};
            
            \node[circle,draw,fill=blue!20] (h51) at (10,0.75) {};
            \node[circle,draw,fill=blue!20] (h53) at (10,-0.75) {};
            \path (h51) -- (h53) node [black, font=\LARGE, midway, sloped, xshift=0.8mm] {$\dots$};
            
            \node[circle,draw,fill=red!20] (output) at (12,0) {};
            
            \foreach \source in {input}
                \foreach \dest in {h11, h13}
                    \draw (\source) -- (\dest);    
            \foreach \source in {h11, h13}
                \foreach \dest in {h21, h23, h25}
                    \draw (\source) -- (\dest);
            \foreach \source in {h21, h23, h25}
                \foreach \dest in {h31, h33, h34, h35, h37}
                    \draw (\source) -- (\dest);
            \foreach \source in {h31, h33, h34, h35, h37}
                \foreach \dest in {h41, h43, h45}
                    \draw (\source) -- (\dest);
            \foreach \source in {h41, h43, h45}
                \foreach \dest in {h51, h53}
                    \draw (\source) -- (\dest);
            \foreach \source in {h51, h53}
                \draw (\source) -- (output);
            
            \coordinate (start) at ($(h11.south west)!0.5!(h51.south east) + (-4.2,-3.0)$);
            \coordinate (end) at ($(h51.south east) + (0,-3.0)$);
            
            \node[below, font=\scriptsize] at (input.south) {Input};
            \node[below, font=\scriptsize, align=center] at ($(h11.south)!0.5!(h13.south) + (-0.3,2.0)$) {64 neurons\\ w/ ReLU};
            \node[below, font=\scriptsize, align=center] at ($(h21.south)!0.5!(h25.south) + (-0.3,2.7)$) {128 neurons\\ w/ tanh};
            \node[below, font=\scriptsize, align=center] at ($(h31.south)!0.5!(h37.south) + (0,3.5)$) {256 neurons\\ w/ tanh};
            \node[below, font=\scriptsize, align=center] at ($(h41.south)!0.5!(h45.south) + (0.3,2.7)$) {128 neurons\\ w/ tanh};
            \node[below, font=\scriptsize, align=center] at ($(h51.south)!0.5!(h53.south) + (0.3,2.0)$) {64 neurons\\ w/ ReLU};
            \node[below, font=\scriptsize, align=center] at (output.south) {Output\\ w/ sigmoid};
        \end{scope}

        \draw[line width=5pt, gray, -latex] (2.5,-3.4) -- (3.5,-3.4);

        \node[align=center] at (5.3,-0.6) {Amplitude-Fidelity\\ Predicted Pairs};
    
        \node[align=center] at (5.3,-3.4) {$(\phi_i, \hat{\chi}_i) = \begin{Bmatrix} (\phi_1, \hat{\chi}_1)\\ (\phi_2, \hat{\chi}_2)\\ (\phi_3, \hat{\chi}_3)\\ (\phi_4, \hat{\chi}_4)\\ \vdots \\ (\phi_n, \hat{\chi}_n) \end{Bmatrix}$};

        \draw[line width=5pt, gray, -latex] (7.1,-3.4) -- (8.1,-3.4);

        \node[align=center] at (10.1,-0.6) {Amplitude That Yields\\ Maximum Fidelity};

        \node[align=center] at (10.0,-3.4) {$(\phi_*, \hat{\chi}_*) = \arg \max\limits_{(\phi_i, \hat{\chi}_i)} \hat{\chi}_i$\\ \\ Desired amplitude: $\textcolor{red}{\boxed{\textcolor{black}{\phi_*}}}$ };
        
    \end{tikzpicture}
    \caption{Schematic for predicting fidelity based on amplitudes: A DNN trained on the amplitude-fidelity database is established. The DNN is then used to predict a large set of new data. Then the amplitude that corresponds to the largest fidelity in the new dataset is selected as the optimum amplitude.}
    \label{predicting_schematic}
  \end{subfigure}
  \caption{Schematic of research process: From data generation to the training of a deep neural network and predicting new amplitude-fidelity data.}
  \label{research_schematic}
\end{figure*}
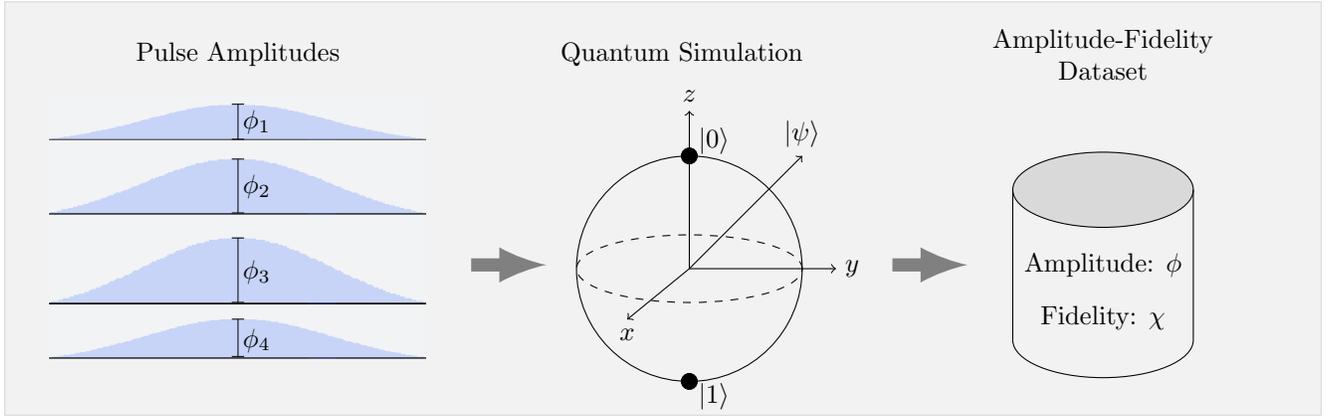
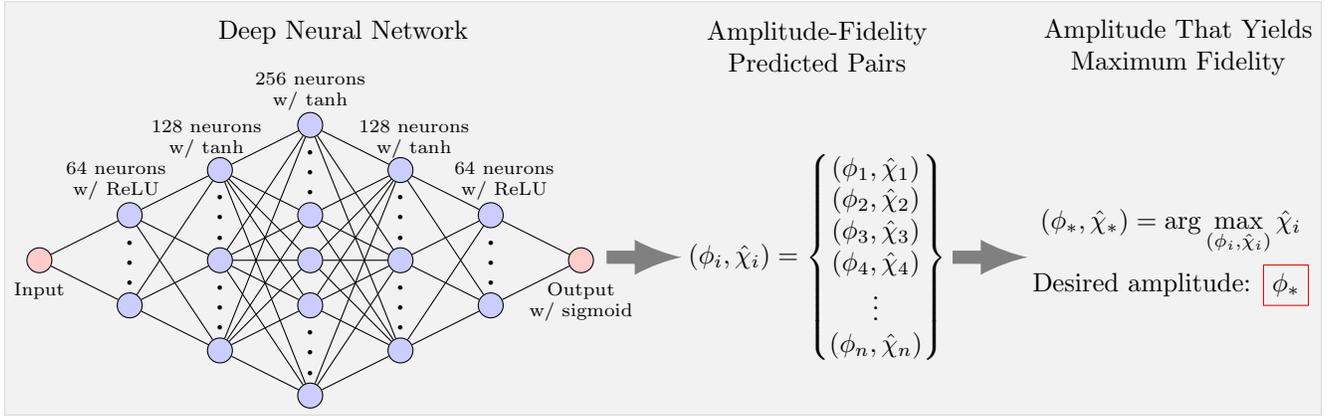

\section{II. BACKGROUND}
\label{background}
\noindent 
In this section, an explanation of foundational qubit gates, namely Hadamard, Pauli-X, and CNOT gates is presented.

\subsection{A. Hadamard Gate}
\label{h_math}
The Hadamard gate, denoted as $H$, is a fundamental qubit gate that introduces superposition between states $|0\rangle$ and $|1\rangle$. Applied to $|0\rangle$ and $|1\rangle$, it yields the states $|\psi\rangle$ as follows:
\begin{equation}
  |\psi\rangle = H|0\rangle = \frac{|0\rangle+|1\rangle}{\sqrt{2}},
  \label{H_applied_zero}
\end{equation}
\begin{equation}
  |\psi\rangle = H|1\rangle = \frac{|0\rangle-|1\rangle}{\sqrt{2}}.
  \label{H_applied_one}
\end{equation}
When a qubit starts in state $|0\rangle$, the resulting state is $|\psi\rangle = H|0\rangle$. The probabilities of measuring $|0\rangle$ and $|1\rangle$ are both $1/2$:
\begin{equation}
    |\langle 0|\psi\rangle|^2 = \left|1/\sqrt{2}\right|^2 = 1/2
\end{equation}
\begin{equation}
    |\langle 1|\psi\rangle|^2 = \left|1/\sqrt{2}\right|^2 = 1/2
\end{equation}
Thus, the Hadamard gate creates an equal superposition of $|0\rangle$ and $|1\rangle$, resulting in a 50\% probability of measuring either state (see Figure~\ref{hist_ideal_h}).

\subsection{B. Pauli-X Gate}
\label{x_math}
The Pauli-X gate, denoted as $X$, is a single-qubit gate in quantum computing, analogous to a NOT gate in classical computing. It flips the qubit state:
\begin{equation}
  X|0\rangle = |1\rangle
\end{equation}
\begin{equation}
  X|1\rangle = |0\rangle
\end{equation} 
Thus for a qubit initially in $|0\rangle$, the $X$ gate results in a 0\% probability of measuring $|0\rangle$ and a 100\% probability of measuring $|1\rangle$ (see Figure~\ref{hist_ideal_x}).

\subsection{C. CNOT Gate}
\label{cnot_maths}
The CNOT (Controlled-NOT) gate is a two-qubit gate which operates on a control qubit $|c\rangle$ and a target qubit $|t\rangle$. 
When the control qubit is $|0\rangle$, the CNOT gate leaves the target qubit unchanged:
\begin{equation}
  |c=0,t\rangle \xrightarrow{CNOT} |c=0,t\rangle
\end{equation}
When the control qubit is $|1\rangle$, the CNOT gate flips the target qubit:
\begin{equation}
  |c=1,t\rangle \xrightarrow{CNOT} |c=0,\neg t\rangle
\end{equation}
Thus if both qubits start in the $|0\rangle$ state:
\begin{equation}
  |c=0,t=0\rangle \xrightarrow{CNOT} |c=0,t=0\rangle
\end{equation}
In this case, the probability of measuring $|00\rangle$ is 100\% (see Figure~\ref{hist_ideal_cnot}).

\section{III. Methods}
\label{sec:methods}
This paper aims to optimise the amplitude of a pulse waveform that operates various qubit gates, such that the fidelity of that gate is as close to ideal as possible. Figure~\ref{research_schematic} shows a block diagram of the approach proposed in this paper. We performed pusle-level simulations of qubit gates to collect a dataset of amplitudes and associated fidelities. The collected dataset is used to train a DNN for predicting fidelities for a given amplitude. A separate DNN is trained for each qubit gate. These DNNs are then used to predict the fidelity depending on the amplitude. The trained DNNs are used to predict fidelity for unforeseen pulse amplitudes, which is a lot less computationally intensive and allows for fine-tuned predictions.

\subsection{A. Data Collection with Qiskit}
\label{qiskit_pulse_sec}

In this paper, Qiskit \cite{qiskit_IBM} is used to perform simulations to collect data mapping amplitudes to fidelities. Qiskit supports multiple backends to simulate quantum circuits at different levels of abstraction. Here, the \textit{FakeValencia} \cite{qiskit_fakevalencia} backend is used to generate pulse waveforms with unique amplitudes. The generated waveforms are used to create pulse schedules for controlling custom implementation of different gates which include Hadamard, Pauli-x and CNOT. 

Our custom implementation of the gates also includes a specialized measurement gate which is used to capture the outcome of the particular gates. The complete circuit is then executed using the \textit{FakeValencia} backend. The flowchart for this iterative process of data generation is shown in Figure~\ref{process_flowchart}.

Each custom implementation is executed using the backend 1024 times per amplitude value. For single-qubit gates, this process is repeated across 400 different amplitude values to generate a dataset. During each execution, the frequency of each qubit state output by the gate is recorded. The probability of each qubit state is then calculated by dividing the count of occurrences by the total number of experiments (1024 per amplitude). These probabilities are used to estimate the fidelity \( (\chi_i) \) between the experimental and ideal gate, for a given amplitude \( (\phi_i) \), using the Bhattacharyya fidelity \cite{nielsen2010quantum}:
\begin{equation}
  \chi_i = \left|\sum_s \sqrt{p_s^e(\phi_i)}\cdot \sqrt{p_s^d}\right|^2,
  \label{eq:fidelity}
\end{equation}
where $\chi_i$ represents the fidelity between the experimental probabilities ($p_s^e(\phi_i)$) and ideal/desired probabilities ($p_s^d$), calculated over each state ($s$) in the distribution of gate states. A value of 1 indicates a perfect match, while lower values signify discrepancies, helping evaluate the performance of the custom qubit gate based on initially defined amplitudes. This process is repeated 400 times for different amplitude values. The resulting combination of amplitude and corresponding fidelities represent the dataset for training deep neural networks.

\begin{Figure}
 \centering
\begin{tikzpicture}[node distance=2cm, scale=0.7, transform shape]
  \draw[gray!30,fill=gray!10] (-5.5,-13.5) rectangle (5.5,0.7);
  \node (start) [startstop] {Define Parameters};
  \node (pro1) [process, below of=start, yshift=0.5cm, minimum width=1.0cm] {Pulse Waveform};
  \node (pro2) [process, below of=pro1, yshift=0.5cm, minimum width=1.0cm] {Qubit Gate};
  \node (pro3) [process, below of=pro2, yshift=0.5cm, minimum width=1.0cm] {Quantum Circuit};
  \node (pro4) [process, below of=pro3, yshift=0.5cm, minimum width=1.0cm] {Pulse Schedule};
  \node (pro5) [process2, right of=pro3, xshift=1.9cm, minimum width=0.5cm] {
      Ideal 
      Circuit
  };
  \node (pro6) [decision, below of=pro4, yshift=-0.5cm] {Run\\ Circuits};
  \node (result1) [io, below of=pro6, xshift=-2cm, yshift=-0.3cm, minimum width=1.5cm] {Realistic States};
  \node (result2) [io, below of=pro6, xshift=2cm, yshift=-0.3cm, minimum width=1.5cm] {Ideal States};
  \node (result3) [startstop, below of=pro6, yshift=-2.3cm, minimum width=2.5cm] {Fidelity};
  \draw [arrow] (start) -- (pro1);
  \draw [arrow] (pro1) -- (pro2);
  \draw [arrow] (pro2) -- (pro3);
  \draw [arrow] (pro3) -- (pro4);
  \draw [arrow] (pro3) -- (pro5);
  \draw [arrow] (pro4) -- (pro6);
  \draw [arrow] (pro5) |- (pro6);
  \draw [arrow] (pro6) -- (result1);
  \draw [arrow] (pro6) -- (result2);
  \draw [arrow] (result1) -- (result3);
  \draw [arrow] (result2) -- (result3);
  \draw [arrow] (result3.west) -- ++(-3.2,0) |- node[pos=0.25, sloped, above] {Iterate} node[pos=0.25, sloped, below] {Simulation} (start);
  \end{tikzpicture}
  
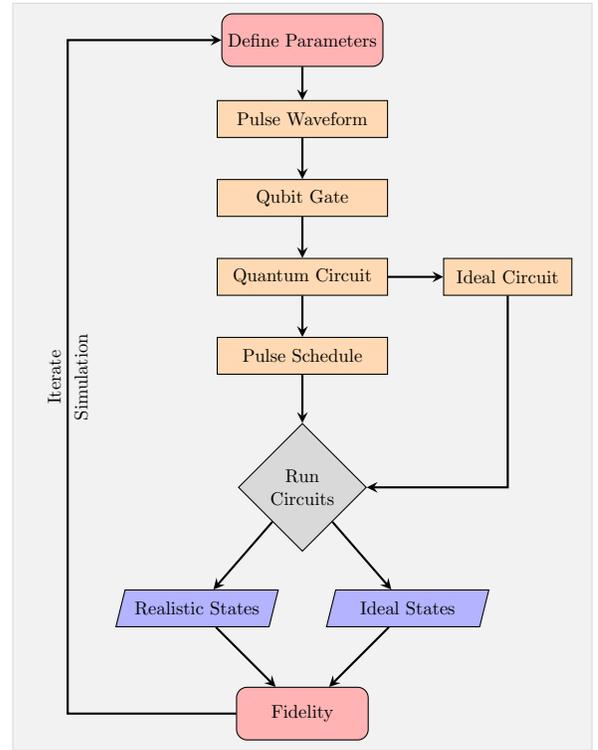
\captionof{figure}{The process of simulating a quantum circuit from pulse-level to find the fidelity in order to optimise the parameters (amplitude).}
  \label{process_flowchart}
\end{Figure}
\noindent

\begin{figure*}[t] 
  \centering
  \begin{tikzpicture}
  \draw[gray!30,fill=gray!10] (-8,-0.7) rectangle (-3.1,3.0);
  \draw[gray!30,fill=gray!10] (-2.8,-0.7) rectangle (2.3,3.0);
  \draw[gray!30,fill=gray!10] (2.7,-0.7) rectangle (7.7,3.0);
  \node [draw=gray!100, inner sep=8pt, line width=0.5pt, rounded corners=5pt] {
  \begin{subfigure}[b]{0.3\textwidth}
    \includegraphics[width=\linewidth]{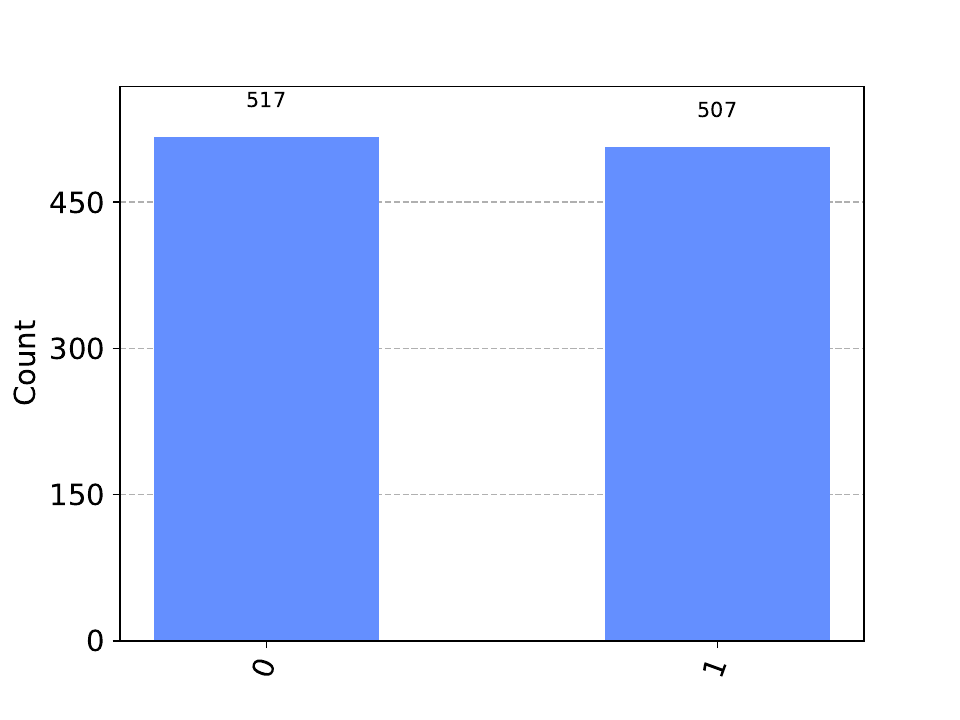}
    \caption{Post-application state distribution of a custom Hadamard gate with amplitude 0.588271, yielding a fidelity of 0.999976. Amplitude prediction derived from a DNN trained on 400 simulations.}
    \label{h_state_distribution}
  \end{subfigure}
  \hfill
  \begin{subfigure}[b]{0.3\textwidth}
    \includegraphics[width=\linewidth]{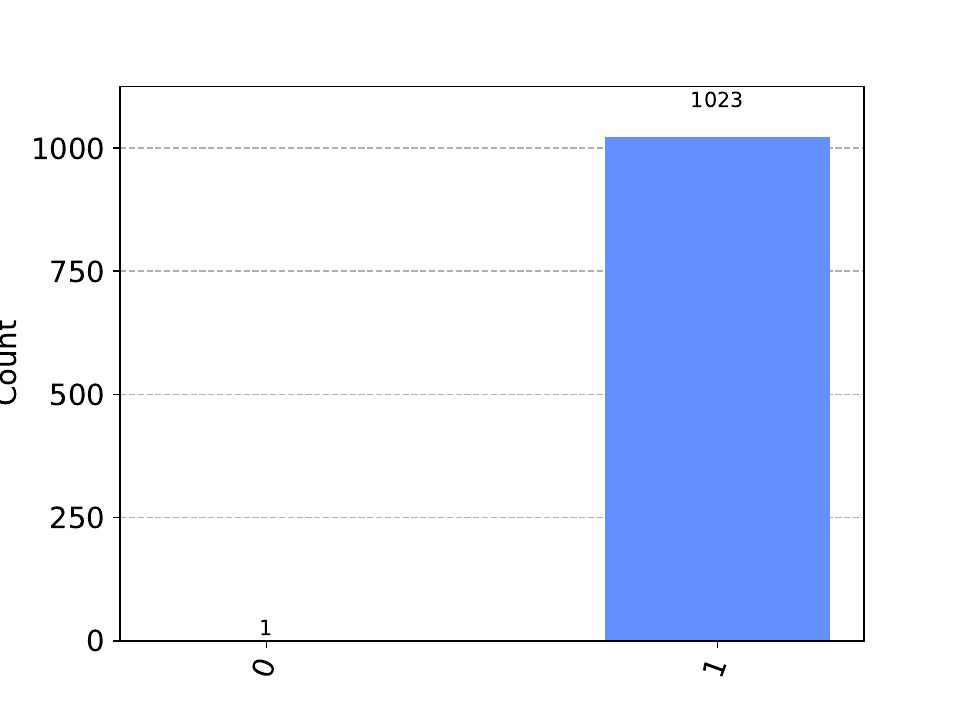}
    \caption{Post-application state distribution of a custom Pauli-X gate with amplitude 0.168764, yielding a fidelity of 0.999923. Amplitude prediction derived from a DNN trained on 400 simulations.}
    \label{x_state_distribution}
  \end{subfigure}
  \hfill
  \begin{subfigure}[b]{0.3\textwidth}
    \includegraphics[width=\linewidth]{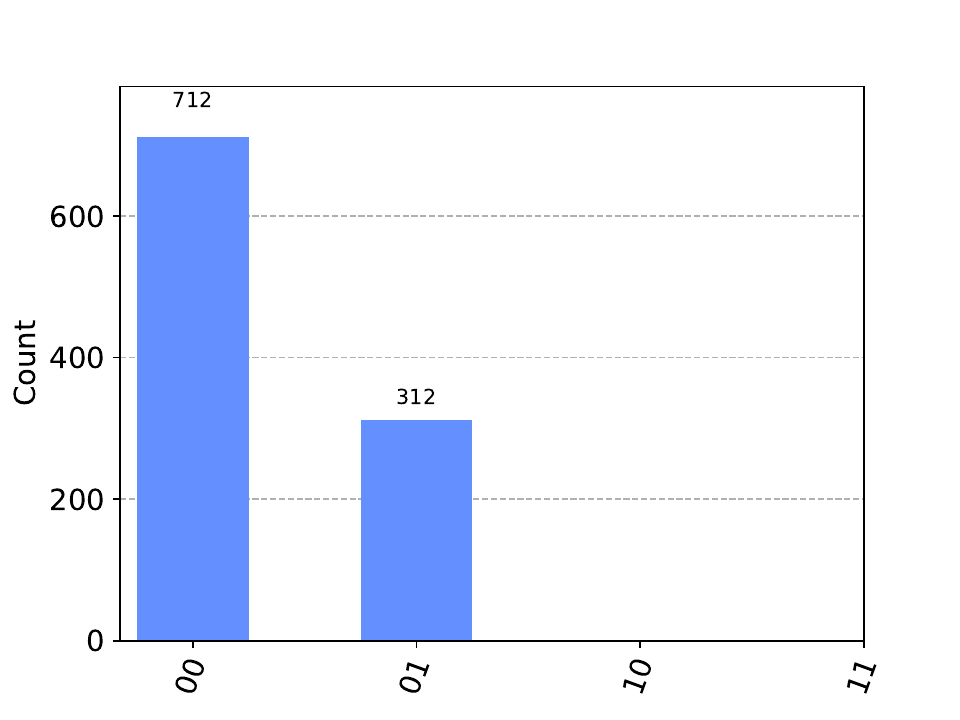}
    \caption{Post-application state distribution of a custom CNOT gate with DRAG amplitudes set to 0.94, 0.27, and 0.72, resulting in a fidelity of 0.695313. Amplitude prediction derived from a DNN trained on 400 simulations.}
    \label{cnot_dis}
  \end{subfigure}
  };
  \end{tikzpicture}
  \caption{The outcome distributions of applying the Hadamard, Pauli-X, and CNOT gates to qubits initialised as $|0\rangle$, with optimised pulse waveforms.}
  \label{outcome_distributions}
\end{figure*}

This process is used to simulate and generate amplitude-fidelity combinations for the Hadamard and Pauli-X gate. The predefined parameter values \cite{parameters_qiskit} (such as duration, phase, and frequency), for the generic Qiskit Hadamard and Pauli-X gate were maintained except for the amplitudes, which were chosen at random from a uniform distribution in the range \( [0, 1] \). These parameters are used to create the pulse waveform for the aforementioned process. As detailed in Section~\hyperref[h_math]{\textcolor{cyan(process)}{IIA}}, perfect fidelity for the Hadamard gate with an initial state of $|0\rangle$ results in a state distribution of $\{|0\rangle: 50\%, |1\rangle: 50\%\}$. Similarly, as detailed in Section~\hyperref[x_math]{\textcolor{cyan(process)}{IIB}}, perfect fidelity for the Pauli-X gate with an initial state of $|0\rangle$ results in a state distribution of $\{|0\rangle: 0\%, |1\rangle: 100\%\}$. However, in practice, the observed state distributions will deviate from these ideal values, resulting in less-than-perfect fidelities. These observed fidelities, along with their initial amplitudes, constitutes the dataset for the Hadamard and Pauli-X gates. 

The same process is also employed for collecting the dataset of amplitude-fidelity combinations for the CNOT gate. The only difference is that three drag pulses \cite{drag_qiskit} are used for simulating the CNOT gate. As laid out in Section~\hyperref[cnot_maths]{\textcolor{cyan(process)}{IIC}}, perfect fidelity for a CNOT gate with qubits initialised in the state $|0\rangle$ results in a state distribution of $\{|00\rangle: 100\%, |01\rangle: 0\%, |10\rangle: 0\%, |11\rangle: 0\%\}$.

Optimising even a single amplitude gate involves exploring a continuous amplitude value space, requiring simulations at fine intervals for the best fidelity. This becomes time-consuming and resource-intensive, especially when extending the approach to various single and multi-qubit gates with more amplitudes and diverse hardware/noise models. In this paper, we employ DNNs to model the relationship between amplitude and fidelities for a given quantum gate using a limited number of simulations. The trained DNN is used to predict the fidelity for a large number of amplitude values to identify the amplitude resulting in highest fidelity.

\subsection{B. Deep Neural Networks for Predicting Fidelity}
\label{general_dnn}
In this section, the DNN to predict fidelity, based on the amplitude of the pulse waveform for qubit gates, is presented. The model is applied to the fundamental single-qubit Hadamard and Pauli-X gates and the multi-qubit CNOT gate, as these gates are key building blocks of more complicated quantum circuits.

Figure~\ref{predicting_schematic} shows the architecture of the DNNs with seven layers used in this paper. While the schematic illustrates a single input node for the DNN, this applies only to single-qubit gates. For the CNOT gate, there are three input nodes, though all other aspects remain the same. The sigmoid function is used as the activation function in the output layer as its output is in the interval \( [0, 1] \) which is same as the fidelity values obtained using Equation~\ref{eq:fidelity}. 

Each amplitude-fidelity combination \( (\phi_i, \chi_i) \) constitutes a single sample for the DNN. The input to the DNN is a given amplitude \( (\phi_i) \) and the predicted fidelity is denoted by \( (\hat{\chi}_i) \). The DNN is trained to minimize the difference between the predicted fidelity ($\hat{\chi}_i$) and the experimental fidelity ($\chi^e_i$) using mean squared error as the loss function \( (\mathcal{L}) \), given as
\begin{equation}
    \mathcal{L} = (\hat{\chi}_i - \chi^e_i)^2 .
\end{equation}
The Adam optimiser \cite{adam_op} with a learning rate of 0.001 is used to update the network parameters. The DNNs reported in this paper are trained for 60 epochs.

\subsection{C. Two Stage Approach to Estimate Pulse Amplitudes}
\label{method_sec_c}
The appropriate amplitude for a given qubit gate is estimated using a two-stage approach using the trained network. In the first stage, amplitudes in the interval of \( [0, 1] \) at increments of \( 1e-3 \) are generated, and the corresponding fidelities are obtained using the trained DNNs, which were trained on data of the scale \( 1e-9 \). The amplitude \( ( \phi_* )\) yielding highest fidelity is identified. In the second stage, amplitudes in the neighborhood of \( \phi_* \), given as \( [\phi_* - \epsilon, \phi_* - \epsilon] \) with increments of \( 1e-6 \) are obtained. Here, \( \epsilon \) is set to \( 1e-3 \). The second stage allows a fine-grained exploration of amplitudes in the neighborhood of the high fidelity amplitude obtained in the first stage.

\section{IV. RESULTS}
\label{result_sec}
In this section, the results of the proposed method for estimating high fidelity amplitudes of pulse waveforms for Hadamard, Pauli-X and CNOT quantum gates are presented. In addition, we also studied the impact of the size of the training data on the fidelities obtained using the proposed method for Hadamard and Pauli-X gates. 

All quantum simulations and models developed in this paper have been conducted on a computer with Intel Core i7-9700 3GHz, NVIDIA GeForce RTX 2080 Ti GPU and 64GB of RAM.

\subsection{A. Evaluation of the Estimated Amplitudes}
The amplitude obtained using the method described in Section~\hyperref[method_sec_c]{\textcolor{cyan(process)}{IIIC}} is simulated in Qiskit to obtain actual fidelities for each quantum gate. Figure \ref{outcome_distributions} shows the final distributions for each output qubit state in Hadamard, Pauli-X and CNOT gates, respectively. The input qubit state is $|0\rangle$ for Hadamard and Pauli-X gates. For CNOT gate, the initial qubit state is $|00\rangle$.

Amplitude values of 0.588271 and 0.168764 are obtained for Hadamard and Pauli-X, respectively. For the CNOT gate, the three amplitudes are 0.94, 0.27, and 0.72. These amplitudes resulted in fidelities of 0.999976, 0.999923 and 0.695313 for the Hadamard, Pauli-X and CNOT gates, respectively. It can be clearly seen from the figure that the proposed method is able to obtain close to optimal fidelities for Hadamard and Pauli-X gates. 
In previously literature \cite{matekole2022methods}, the L-BFGS-B has obtained probability distributions of $\{|0\rangle: 0.565, |1\rangle: 0.435\}$ and $\{|0\rangle: 0.127, |1\rangle: 0.873\}$ for the Hadamard and Pauli-X gate, respectively. These distributions yield fidelities of 0.99575 and 0.87300 for their respective gates. It is evident from these fidelities that the method demonstrated within this paper has achieved significantly high fidelities for the Hadamard and Pauli-X gates, demonstrating the effectiveness of our optimisation approach.

Due to the complexity of optimising two-qubit gates, the amplitudes obtained for the CNOT gate, did not yield high fidelities. A state distribution of $\{|00\rangle: 712, |01\rangle: 312, |10\rangle: 0, |11\rangle: 0\}$ is achieved for the amplitude obtained using the proposed method, resulting in a fidelity of 0.695313 (see Figure~\ref{cnot_dis}). The challenge stems from the intricacies of two-qubit gates, involving entangled states and increased computational demands. For CNOT gates, phase becomes a significant parameter for optimisation. 
While the phase parameter was excluded for single-qubit gates due to its negligible effect, it plays a crucial role in multi-qubit gates, influencing qubit synchronization.

It may be noted that there are multiple amplitudes resulting in high fidelities for Hadamard and Pauli-X gates. This is due to the fact that the relationship between fidelity and the amplitude in a pulse waveform controlling a single-qubit gate is sinusoidal due to the quantum mechanical phenomenon of Rabi oscillations \cite{nielsen2001quantum}.

\subsection{B. Impact of the Size of Training Data}
Simulating quantum circuits is computationally expensive. Therefore, a computationally efficient method to estimate amplitudes yielding high fidelities should require fewer simulations. The training data for the proposed method is obtained through quantum simulation using Qiskit. To understand the computationally efficiency of the proposed method, we studied the impact of size of the training data on its performance in terms of the final fidelities obtained. For this purpose, DNNs are trained using different sizes of the training data and the amplitude resulting in highest fidelity is obtained using the method described in the Section~\hyperref[method_sec_c]{\textcolor{cyan(process)}{IIIC}}. A quantum circuit using the identified amplitude is simulated in Qiskit to obtain actual fidelity.

Figure~\ref{gate_fidelity_plot} illustrates the relationship between dataset size and corresponding fidelity values for both the Hadamard and Pauli-X gates. For the Hadamard gate, the results indicate diminishing returns after 175 simulations, with an amplitude of 0.415085 achieving a fidelity of 0.990225. Similarly, for the Pauli-X gate, diminishing returns are observed after approximately 50 simulations, with a \blockcomment{highest predicted}fidelity of 0.977539 achieved at an amplitude of 0.519873. When the training dataset includes the total 400 simulations, the achieved fidelities are 0.999976 for the Hadamard gate and 0.999923 for the Pauli-X gate.

\begin{Figure}
  \centering
  \begin{tikzpicture}
    \draw[gray!30,fill=gray!10] (-4,-4.7) rectangle (4,1.3);
    \node at (0, -1.75) {\includegraphics[width=8cm]{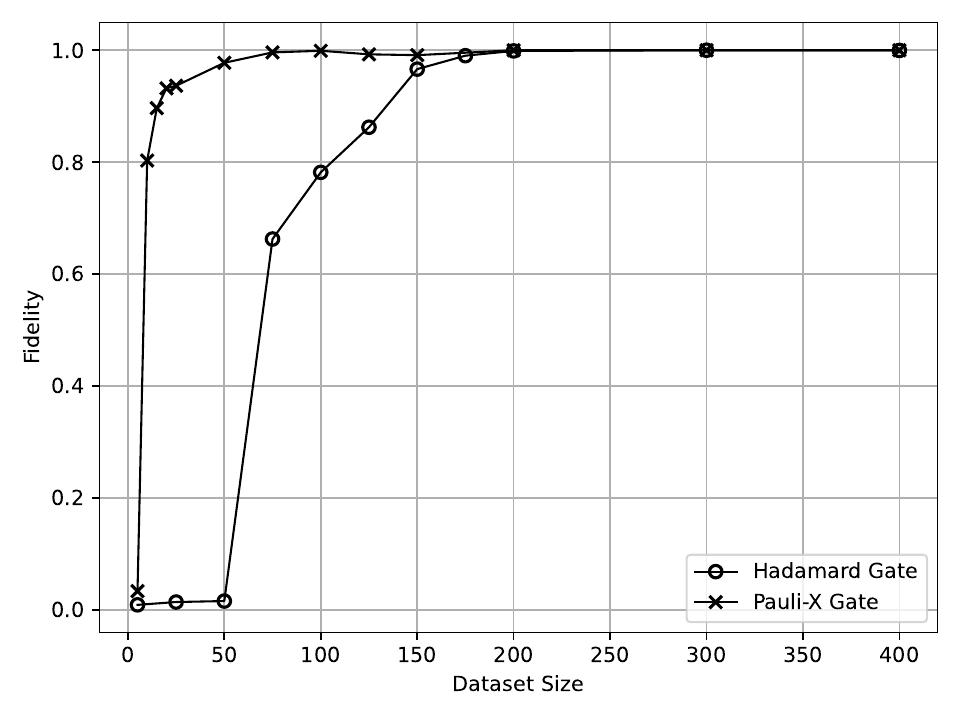}};
  \end{tikzpicture}
  \captionof{figure}{Dataset size vs fidelity for both the Hadamard and Pauli-X gate.}
  \label{gate_fidelity_plot}
\end{Figure}

We also studied the relationship between dataset size and training loss curves by examining the Spearman's correlation coefficient between the training and validation curves. This analysis helps determine how well the model generalizes during training. A strong positive correlation between the training and validation curves indicates that the model's performance on the training data is predictive of its performance on the validation data.

Figure~\ref{gate_spearman_plot} illustrates the relationship between dataset size and the corresponding Spearman's correlation coefficient for both the Hadamard and Pauli-X gates. For the Hadamard gate, diminishing returns are observed after 300 simulations, achieving a Spearman's correlation coefficient of 0.9820. Similarly, for the Pauli-X gate, diminishing returns occur after approximately 75 simulations, with a Spearman's correlation coefficient of 0.9915.

\begin{Figure}
  \centering
  \begin{tikzpicture}
    \draw[gray!30,fill=gray!10] (-4,-4.7) rectangle (4,1.3);
    \node at (0, -1.75) {\includegraphics[width=8cm]{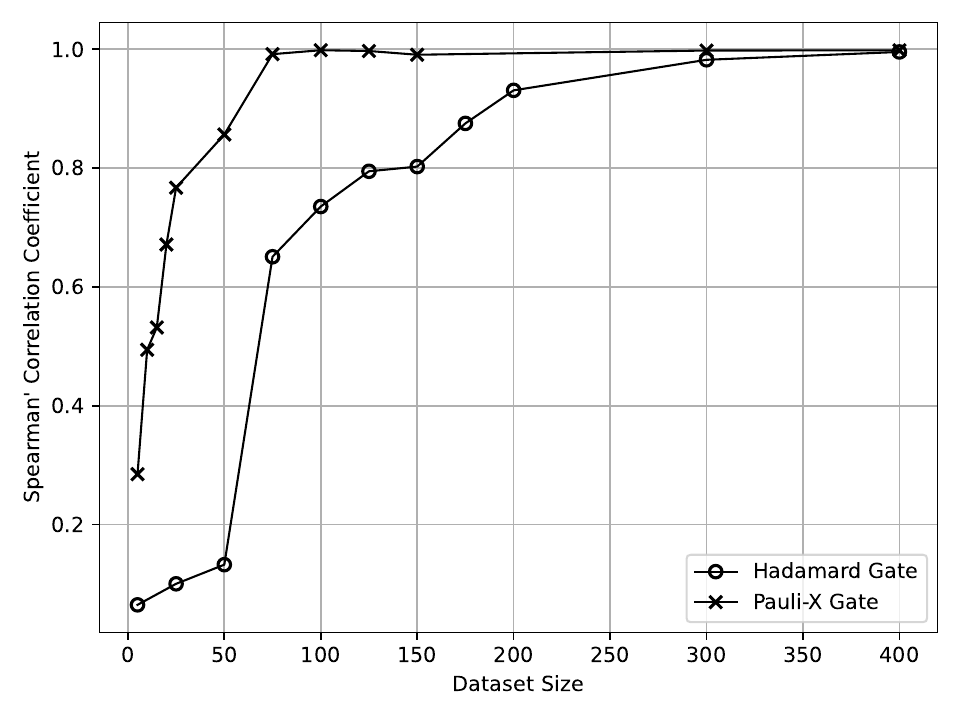}};
  \end{tikzpicture}
  \captionof{figure}{Dataset vs Spearmans's correlation coefficient for both the Hadamard and Pauli-X gate.}
  \label{gate_spearman_plot}
\end{Figure}

\section{V. CONCLUSION AND DISCUSSION}
In this paper, we have presented a DNN-based approach to obtain high fidelity simulations of qubit gates by optimising the parameters associated with pulse scheduling. These results demonstrate DNNs can be used to optimise the amplitudes of pulse schedules for single-qubit gates.

The findings have several implications. Firstly, they enable more efficient and accurate quantum computations, enhancing quantum algorithm performance and potentially solving previously intractable problems. The methodology is extendable to multi-qubit gates, such as the CNOT gate; however, extending this optimisation to the CNOT gate introduces complexities related to quantum entanglement and multi-qubit control. Addressing these challenges requires a deeper understanding of quantum control theory and may involve adapting the current deep learning framework.

Future research could focus on quantum control theory for multi-qubit gates. Once multi-qubit gates are optimised, this research can extend to entire quantum circuits. Optimising circuits presents new challenges, including the need for comprehensive information beyond individual gate amplitudes and phases, such as gate sequences and adjacency matrices to describe qubit connectivity accurately. A similar DNN model can be adapted to handle these new challenges, accepting input data that includes gate sequences, adjacency matrices for representing connectivity between qubits, and other circuit-specific information. Thus, with increasing complexity of gates and circuits, the proposed data-driven approach could overcome the shortcomings of the classical methods by modelling noise in quantum systems.

In terms of computational complexity, the proposed approach significantly reduces computational resources and time, allowing continuous refinement of the prediction space to find increasingly precise solutions. On our computer, a quantum simulation of the Hadamard gate took 3.0 seconds, whereas the DNN prediction required only 0.035 seconds.  This represents an almost hundredfold reduction in computation time compared to the simulation. While, these numbers do depend on other processes using the computation resources, they do reflect the potential benefits of the proposed approach. In quantum mechanics, where small variations can lead to drastically different outcomes, this refinement is crucial.

\bibliographystyle{unsrt}
\bibliography{bibliography}
\end{multicols}

\end{document}